\documentclass[aps,superscriptaddress,showpacs,nofootinbib,eqsecnum,prd,twocolumn]{revtex4}

\usepackage{epsfig}  
\input{epsf}
\def\be{\begin{equation}} 
\def\ee{\end{equation}} 
\def\bea{\begin{eqnarray}}
\def\eea{\end{eqnarray}}

\def\beq{\begin{equation}}
\def\eeq{\end{equation}}
\def\beqa{\begin{eqnarray}}
\def\eeqa{\end{eqnarray}}

\begin{document}

\title{The Importance  of Asymptotic Freedom for the Pseudocritical Temperature 
in Magnetized Quark Matter}

\author{R.L.S. Farias} \email{ricardofarias@ufsj.edu.br}
\affiliation{Departamento de Ci\^{e}ncias Naturais, Universidade Federal de
  S\~ao Jo\~ao Del Rei, 36301-000, S\~ao Jo\~ao Del Rei, MG, Brazil}
\affiliation{Departamento de F\'isica, Universidade Federal de
  Santa Maria, 97105-900, Santa Maria, RS, Brazil}
  
\author{K.P. Gomes} \email{karinaponcianogomes@gmail.com} 
\affiliation{Departamento de Ci\^{e}ncias Naturais, Universidade Federal de
  S\~ao Jo\~ao Del Rei, 36301-000, S\~ao Jo\~ao Del Rei, MG, Brazil}

\author{G. Krein} \email{gkrein@ift.unesp.br}
\affiliation{Instituto de F\'{\i}sica Te\'orica, Universidade Estadual
  Paulista,  Rua Dr. Bento Teobaldo Ferraz, 271 - Bloco II, 
  01140-070 S\~ao Paulo, SP, Brazil}

\author{M.B. Pinto} \email{marcus.benghi@ufsc.br}
\affiliation{Departamento de F\'{\i}sica, Universidade Federal de Santa
  Catarina, 88040-900 Florian\'{o}polis, Santa Catarina, Brazil}

\begin{abstract}
Although asymptotic freedom is an essential feature of QCD, it is absent in effective 
chiral quark models like the Nambu--Jona-Lasinio and linear sigma models. In this work we 
advocate that asymptotic freedom plays a key role in the recently observed discrepancies 
between results of lattice QCD simulations and quark models regarding the behavior of the 
pseudocritical temperature $T_{\rm pc}$ for chiral symmetry restoration in the presence of 
a magnetic field $B$. We show that the lattice predictions that $T_{\rm pc}$ decreases with 
$B$ can be reproduced within the Nambu--Jona-Lasinio model if the coupling constant 
$G$ of the model decreases with $B$ and the temperature. Without aiming at numerical 
precision, we support our claim by considering a simple ansatz for $G$ that mimics 
the asymptotic freedom behavior of the QCD coupling constant $1/\alpha_s \sim 
\ln (eB/\Lambda^2_{ QCD})$ for large values of $B$. 
\end{abstract}

\pacs{21.65.Qr,25.75.Nq,11.30.Rd,11.10.Wx,12.39.-x}

\maketitle

\section{Introduction}

The  investigation of the effects produced by a  magnetic field in the phase 
diagram of strongly interacting matter became a subject of great interest in recent 
years. The recent motivation stems mainly from the fact that strong magnetic fields may 
be produced in non central heavy ion collisions~\cite{kharzeev1,kharzeev2,kharzeev3} 
- for an updated discussion, see  Ref.~\cite{tuchin}. Strong magnetic fields are 
also present in magnetars~\cite{magnetars1,magnetars2} and might have played an 
important role in the physics of the early universe~\cite{universe}. At vanishing 
baryon density and magnetic field, lattice QCD simulations~\cite{Aoki1,Aoki2} 
predict that there is a crossover transition at a pseudo critical temperature 
$T_{\rm pc}$. More recent lattice simulations~\cite{earlylattice,preussker,lattice1,lattice2} 
indicate that such a crossover persists in the presence of a magnetic field 
and, in addition, confirm predictions made a few years ago~\cite{KlevLemm,Gus,ShushSmilga}  
of the phenomenon of magnetic catalysis (MC), in that the quark condensate is
enhanced in the presence of the magnetic field at vanishing temperatures. These 
lattice results also lend support to calculations using effective models, e.g. those 
of Refs.~\cite{eduardo1,eduardo,prd,andersenBsigma,prd12}. There are, however, marked 
disagreements between lattice results and model calculations regarding the dependence 
of $T_{\rm pc}$ on the strength $B$ of the magnetic field. Specifically, the lattice results 
of Refs.~\cite{lattice1,lattice2}, performed with $2+1$ quark flavors and physical pion 
mass values, predict an inverse catalysis, in that $T_{\rm pc}$ decreases with $B$, 
while effective models predict an increase of $T_{\rm pc}$ with $B$ - an exception is a 
calculation~\cite{mit} based on the MIT bag model which, on the other hand, misses the 
crossover nature of the transition. Earlier model evaluations have been mainly performed 
considering  the two flavor Nambu--Jona-Lasinio model (NJL) \cite {prcsu2} and  the linear 
sigma model~\cite {eduardo1} within the mean field approximation (MFA). Subsequently, 
these evaluations have been considered at more sophisticated levels by coupling degrees 
of freedom related to the Polyakov loop (PNJL models)~\cite {eduardo} and including 
strangeness~\cite {prd}. Calculations have also performed beyond the MFA using the 
functional renormalization group~\cite{skokov,newandersen}. Nevertheless, 
despite those refinements in model calculations, no qualitative changes in the 
behavior of $T_{\rm pc}$ with $B$ have been observed. 

Possible explanations for the disagreement concerning the dependence of $T_{\rm pc}$ on
$B$ have been recently given in Refs.~\cite{prl,kojo,endrodi, leticiaLN}. A~particularly insightful
discussion is the one of Ref.~\cite{endrodi}: the authors of this reference argue that 
the inverse catalysis is the result of the back-reaction of the gluons due to the coupling
of the magnetic field to the sea quarks. Such a back-reaction is naturally implemented in the 
so-called entangled PNJL model~\cite{EPNJL1,EPNJL2}, in that the four-quark coupling of the NJL 
model is made dependent on the Polyakov loop. Although such a model is also unable~\cite{EPNJL3} 
to produce inverse catalysis, a very recent study~\cite{debora} has shown that if the lattice 
data~\cite{lattice2} are fitted by making the pure-gauge critical temperature $T_0$, a parameter 
of PNJL models, depend on $B$, the model gives inverse catalysis. A similar recent study~\cite{bruno} 
in the context of a quark-meson (QM) model has concluded that the only way to get inverse catalysis 
would be to let the Yukawa quark-meson coupling increase with $B$, but no qualitative
agreement with lattice simulations for the phase diagram $T_{\rm pc}(B)$ is obtained. Moreover,
the situation does not improve when this model is extended with the Polyakov loop (PQM)
and the model parameter $T_0$ is made $B$ dependent like in Ref.~\cite{debora}.

From this discussion, it is apparent that the failure of effective models  in
providing inverse catalysis can be traced to a more fundamental level by recalling that 
the couplings in those models do not run with the magnetic field, in contrast with QCD 
whose coupling, as Miransky and  Shovkovy have shown~\cite {miransky}, decreases for large 
$B$ in a manifestation of the asymptotic freedom phenomenon. Therefore, in the present work 
we make use of an ansatz that makes $G$ a running coupling with $B$ and $T$, very much like 
the strong coupling runs in QCD. As a result, we will show that at $T=0$ the model realizes 
magnetic catalysis, and at $T \neq 0$, it realizes inverse catalysis, in qualitative agreement 
with the lattice simulations of Refs.~\cite{lattice1,lattice2}. For zero magnetic field, the 
idea of using an ansatz to mimic high temperature lattice results for the effective QCD 
coupling within the NJL model is not new, it was implemented in Ref.~\cite{bernard} to study 
the decoupling of the pion at high temperatures. A similar ansatz was used very recently
in Ref.~\cite{Qin:2010pc} in a study of the quark spectral density in a strongly-coupled 
quark-gluon plasma in the context of the QCD Dyson-Schwinger equations at finite temperature 
and density -- for a review, see Ref.~\cite{Roberts:2000aa}. 

Our motivation is easily understood by recalling that in the NJL model the quark mass 
gap and the quark condensate are related by $M \sim - G \langle {\bar \psi}_f \psi_f \rangle$,
where  $\langle {\bar \psi}_ f \psi_f \rangle$ grows with $B$, as a manifestation of MC. When
$G$ does not run with $B$ and $T$, the  quark mass gap and $T_{\rm pc}$ also grow with $B$. 
One way to circumvent this problem has been recently suggested by Kojo and 
Su~\cite{kojo} who consider an effective interaction with infrared enhancement 
and ultraviolet suppression. Our proposal is that basically the same effect can be obtained 
by assuming that the NJL quark-quark coupling decreases with $B$ and $T$, as does the QCD 
coupling. As a result, the NJL model results turn out to be in qualitative agreement with the 
recent lattice results; the condensate grows in accordance with MC and the pseudocritical 
temperature for the chiral transition decreases with $B$. Although this agreement comes out 
in a particular model implementation, it lends support to to the suggestion that the inverse
catalysis is the result of the back reaction of the sea quarks to the magnetic 
field~\cite{endrodi}. 

On a wider perspective, the motivation for tuning highly simplified models of the 
NJL type to reproduce the decreasing of $T_{pc}$ with $B$ obtained from much 
more fundamental non-perturbative lattice QCD calculations comes within the same 
spirit as in previous studies using such models. The situation here is similar to the 
calculation of the hadron masses using lattice QCD simulations: while presently correct 
numbers for the masses are generated by the simulations, there is no information on 
the physical mechanisms that generate those numbers from gluon and (essentially 
massless) quark fields. In this respect, the NJL model, in particular, provides physical 
insight into the mechanism behind mass generation via dynamical chiral symmetry breaking. 
In the present context, one is able to understand the decreasing of $T_{pc}$ with $B$ 
by using a magnetic field-induced running of $G$, which is interpreted as being due to 
the indirect coupling of $B$ to the gluons whose physics is similar to that of asymptotic 
freedom. 

The paper is organized as follows. In the next section we discuss the quark condensate
considering the two flavor NJL model for hot and magnetized quark matter in the MFA. The
running of the coupling $G$ is introduced in the same section. Then, in Sec.~\ref{results}, 
we discuss our numerical results while our conclusions and perspectives are presented in 
Sec.~\ref{conclusions}. 
 
\section{Running Coupling in the Magnetized NJL Model}
\label{njlTB}

The standard two flavor NJL model is defined by a fermionic Lagrangian density  
given by~\cite{njl}
\begin{equation}
\mathcal{L}_{\rm NJL}={\bar \psi}\left( i{\partial \hbox{$\!\!\!/$}}-m\right) \psi
+ G\left[ ({\bar \psi}\psi)^{2} 
- ({\bar{\psi}} \gamma _{5}{\vec{\tau}}\psi)^{2}
\right] ,
\label{njl2}
\end{equation}
\noindent
where $\psi$ represents a flavor iso-doublet ($u$ and $d$ flavors) and $N_{c}$-plet 
of quark fields, while $\vec{\tau}$ are isospin Pauli matrices -- a sum over flavor and 
color degrees of freedom is implicit. The Lagrangian density in Eq.~(\ref{njl2}) is 
invariant under (global) $U(2)_{\rm f}\times SU(N_{c})$ and, when $m=0$, the theory 
is also invariant under chiral $SU(2)_{L}\times SU(2)_{R}$. Within the NJL model a 
sharp cut off $\Lambda$ is generally used as an ultraviolet regulator and, since 
the model is nonrenormalizable, one has to fix $\Lambda$ to physical quantities. 
The phenomenological values of quantities such as the  pion mass $m_{\pi}$, 
the pion decay constant $f_{\pi}$, and the quark condensate $ \langle {\bar \psi}_f 
\psi_f \rangle$ are used to fix  $G$, $\Lambda$, and $m$.  Here, we choose the set 
$\Lambda=650\,{\rm MeV}$ and $G=5.022\, {\rm GeV}^{-2}$ with $m=5.5\,{\rm MeV}$ in 
order to reproduce $f_\pi=93\,{\rm MeV}$, $m_\pi= 140\,{\rm MeV}$, and 
$\langle {\bar \psi}_f \psi_f \rangle^{1/3}=-250 \, {\rm MeV}$ in the vacuum.  

The expressions for the thermodynamic potential of the model in the MFA are
well documented in the literature, see e.g. Refs.~\cite {buballa,prcsu2} 
-- for results beyond the MFA, see Ref.~\cite {prc1}. Thus, we refrain 
from repeating them here and start with the expression for the gap 
equation~\cite {prd12}
\begin{equation}
M_f = m_f - 2G \sum_f\langle {\bar \psi}_f \psi_f \rangle ,
\end{equation}
where $\langle {\bar \psi}_f \psi_f \rangle$ represents the quark condensate
of flavor~$f$
\begin{widetext}
\begin{eqnarray}
\langle \bar \psi_f\psi_f \rangle &=& - \frac{ N_c  M_f}{2\pi^2} \left\{
\Lambda\sqrt {\Lambda^2 + M_f^2} - \frac{M_f^2}{2}
\ln \left[ \frac{(\Lambda+ \sqrt {\Lambda^2 + {M_f^2}})^2}{{M_f}^2} \right] \right\} 
\nonumber \\
&-&\frac{N_c M_f}{2\pi^2}  |q_f| B\bigg\{ \ln [\Gamma(x_f)] 
-\frac {1}{2} \ln (2\pi) +x_f  - \frac{1}{2} \left ( 2 x_f-1 \right )\ln(x_f) 
\biggr\} \nonumber \\
&+& \frac{N_c M_f}{2\pi^2} \sum_{k=0}^{\infty} \alpha_k |q_f|B \int_{-\infty}^{\infty} 
\frac{dp_z}{E_{p,k}(B)}\left\{ \frac{1}{e^{[E_{p,k}(B)]/T}+1} \right\} ,
\end{eqnarray}
\end{widetext}
where $E_{p,\,k}(B)=\sqrt{p_z^2+2k|q_f|B +M_f^2}$, $x_f=M_f^2/(2|q_f|B)$, and
$\alpha_k = 2 - \delta_{0k}$. In addition, $|q_f|$ is 
the absolute value of the quark electric charge; $|q_u|= 2e/3$, $|q_d| = e/3$, with 
$e = 1/\sqrt{137}$ representing the electron charge -- we use Gaussian natural units 
where $1\, {\rm GeV}^2 = 1.44 \times 10^{19} \,{\it G}$. Note also that here we have taken 
the chemical equilibrium condition by setting $\mu_u=\mu_d=\mu$. Details of the 
manipulations leading to this equation can be found in the appendix of Ref.~\cite {prcsu2}. 
Note that the condensates for the flavors $u$ and $d$ are different due to their
different electric charges. Remark also that in principle one should have two coupled gap 
equations for the two distinct flavors: $M_u = m_{u} - 2G(\langle {\bar \psi}_u \psi_u \rangle 
+ \langle {\bar \psi}_d \psi_d \rangle)$ and  $M_d = m_{d} - 2G(\langle {\bar \psi }_u \psi_u 
\rangle + \langle {\bar \psi}_d \psi_d \rangle)$. However, in the two flavor case, the 
different condensates contribute to $M_u$ and $M_d$ in a symmetric way and, since $m_u=m_d=m$, 
one can write $M_u = M_d=M$. 

Ref.~\cite{lattice2} presents results for $(\Sigma_u + \Sigma_d)/2$ and
$\Sigma_u - \Sigma_d$, where $\Sigma_f = \Sigma_f(B,T)$ is defined as
 \begin{equation}
\Sigma_{f}(B,T) = \frac{2m_{f}}{m_\pi^2 f_\pi^2}\left[\langle \bar \psi_f\psi_f \rangle 
- \langle \bar \psi_f\psi_f \rangle_0\right]+1 ,
\label{sigmalatt}
 \end{equation}
with $ \langle \bar \psi_f\psi_f \rangle_0$ being the quark condensate at $T=0$ and $B=0$.

Let us recall the important result by Miransky and  Shovkovy~\cite{miransky} that, for 
sufficiently strong magnetic fields, $eB \gg \Lambda^2_{QCD}$, the leading order running 
of the QCD coupling constant $\alpha_s$ is given by
\begin{equation}
\frac{1}{\alpha_s} \sim  b\, \ln\frac{eB}{\Lambda_{QCD}^2} ,
\label{run_QCD}
\end{equation}
where $b = (11N_c - 2N_f)/12\pi$, and the energy scale $\sqrt{eB}$ is fixed up 
to a factor of order~1. Motivated by this result, we propose for the NJL coupling, at $T=0$,
the interpolating formula  
\begin{eqnarray}
G(B) =\frac{G_0}{1+ \alpha \ln \left(1 + \beta \, \frac{eB}{\Lambda^2_{QCD}}\right)},
\label{run_G}
\end{eqnarray}
with $G_0 = 5.022$~{\rm{GeV}$^{-2}$}, which is the value of the coupling at $B=0$.
The free parameters $\alpha$ and $\beta$ are fixed to obtain a reasonable 
description of the lattice average $(\Sigma_u + \Sigma_d)/2$ for $T=0$. In principle, 
for the values of B presently considered in lattice simulations, there is no 
reason for $G(B)$ have a logarithm running like the large-B running of $\alpha_s(B)$.
The proposed parametrization for the running of $G$ with $B$ is motivated, primarily, 
by the assumption that similar physics that drives asymptotic freedom is also 
responsible for the decrease of the effective coupling with the magnetic field; 
namely, the back reaction of gluons via the coupling of the magnetic field on 
the sea quarks. In addition, although the precise $B$ dependence of the effective 
coupling for values of eB not much larger that $\Lambda^2_{QCD}$ is not known, the 
proposed parametrization involves a minimum number of fitting parameters, 
extrapolates smoothly to the large-B running of the quark-gluon running in QCD, 
and can be tested when larger values of $B$ will be used in lattice simulations. 
As we will show in the following, only a decrease (not necessarily in a logarithmic 
way) of $G$ with $B$ can account for the recent lattice results at $T = 0$ . 

At high temperatures, $\alpha_s$ also runs as the inverse of $\ln (T/\Lambda_{QCD})$. 
However, the values of $T$ used in the lattice simulations of Refs.~\cite{lattice1,lattice2}, 
$T \leq \Lambda_{QCD}$, are not high enough to justify the use of such a running for $G$. 
Since the explicit form in which $\alpha_s$ runs with $B$ {\em and} $T$ is presently not 
known, we consider a $T$ dependence for $G$ as 
\begin{equation}
G(B,T) = G(B) \, \left(1 - \gamma\, \frac{|eB|}{\Lambda^2_{QCD}}\,
\frac{T}{\Lambda_{QCD}}\right),
\label{GT}
\end{equation}
where $\gamma$ is fixed to obtain a reasonable description of the temperature dependence 
of the lattice average $(\Sigma_u + \Sigma_d)/2$ at the highest temperatures. Notice
that terms proportional to $T$, $T^2$, $(BT)^ 2$, $\cdots$ could be considered, however 
it turned out that they are not needed to obtain a reasonable fit to the lattice 
data for $(\Sigma_u + \Sigma_d)/2$.  It should be clear that for the values of $B$ used 
in the lattice simulations, any function $G(B)$ that gives an effective coupling that 
decreases with $B$ can fit the lattice data for $(\Sigma_u + \Sigma_d)/2$ for the magnetic 
field values  used in the simulations, if a sufficient number of free parameters are used.

\section{Numerical Results} 
\label{results}

Before presenting our results, we note that Refs.~\cite{lattice1,lattice2} used 
$m_u=m_d=5.5\,\rm{MeV}$, $m_{\pi}=135$ {\rm{MeV}} and $f_\pi=86$ {\rm{MeV} in the 
multiplicative factor $m_{f}/m_\pi^2 f_\pi^2$ in Eq.~(\ref{sigmalatt}) to make 
$\Sigma_f(B,T)$ dimensionless. The fact that these values differ from the ones we 
use is of no significance for comparison purposes since they only set a general scale. 
We also consider $\Lambda_{QCD}=200\,{\rm MeV}$.  

Fig.~\ref{fodormed} displays the condensate average $(\Sigma_u + \Sigma_d)/2$. We obtain a
good fit of the data at $T=0$ with $\alpha =2$ and $\beta = 0.000327$ in Eq.~(\ref{run_G}).
Fig.~\ref{fodormed} reveals that the magnetic catalysis is naturally reproduced at $T=0$. 
In addition, the $T$ dependence is also reasonably well reproduced -- here we used 
$\gamma = 0.0175$ in Eq.~(\ref{GT}). Having fixed our parameters, we proceed in the 
analyses of other quantities. 

\begin{figure}[htb]
\centerline{ \epsfig{file=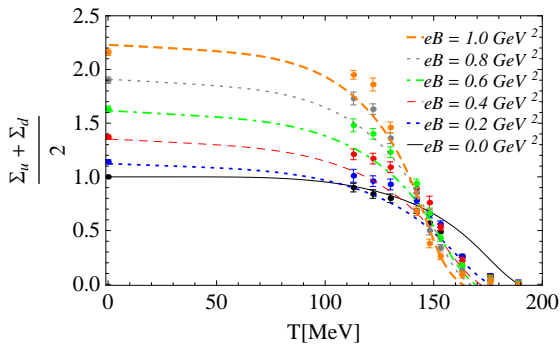,width=0.85\linewidth,angle=0}}
\caption{(color online) The condensate average as a function of the temperature. 
Data points are from the lattice simulations of Ref.~\cite{lattice2}. }
\label{fodormed}
\end{figure}
 
The difference $\Sigma_u - \Sigma_d$ is displayed in Fig.~\ref {fodordiff}. Although
there is a small deviation from the lattice results for the highest values of $B$, the 
overall agreement is quite impressive, given the simplicity of the model.  

\begin{figure}[htb]
\centerline{ \epsfig{file=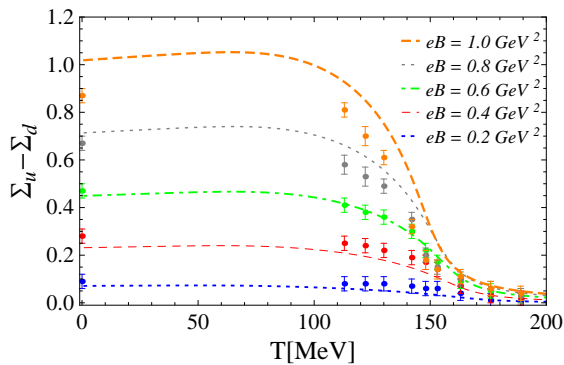,width=0.85\linewidth,angle=0}}
\caption{(color online) The condensate difference as a function of the temperature. 
Data points are from the lattice simulations of Ref.~\cite{lattice2}. }
\label{fodordiff}
\end{figure}

In this work we consider the physical point with  nonzero current quark masses so that, 
at high temperatures, the model displays a crossover where chiral symmetry is partially 
restored. In this case, one can only establish a pseudocritical temperature which 
depends on the observable used to define it. Here, we use the location of the peaks 
for the vacuum normalized quark condensates, where the thermal susceptibilities are 
given by:
\begin{equation}
 \chi_T=-m_{\pi}\frac{\partial \sigma}{\partial T} \,\,,
\label{chiT}
\end{equation}
with $\sigma$ being defined by
\begin{equation}
\sigma=\frac{\langle\bar{\psi}_u\psi_u\rangle(B,T)+\langle\bar{\psi}_d\psi_d\rangle(B,T)}
{\langle\bar{\psi}_u\psi_u\rangle(B,0)
+\langle\bar{\psi}_d\psi_d\rangle(B,0)}\,\,.
\end{equation}
 
In Fig.~\ref{suscep} we plot the thermal susceptibility defined by Eq.~(\ref{chiT}) as a 
function of the temperature for different values of the magnetic field.  The figure clearly 
indicates the decrease of $T_{\rm pc}$ for increasing values of the magnetic field. Again, 
the effects of asymptotic freedom seem to be a rather important feature to conciliate 
results obtained with the NJL model and lattice simulations. 

\begin{figure}[htb]
\centerline{ \epsfig{file=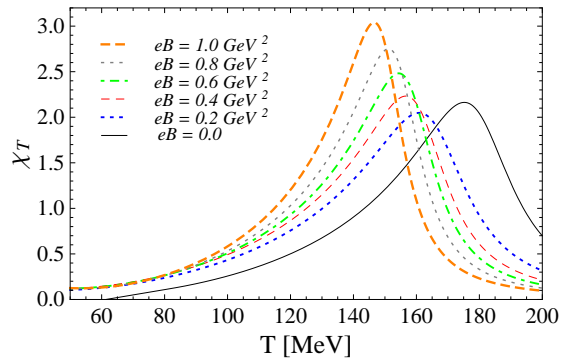,width=0.85\linewidth,angle=0}}
\caption{(color online). The normalized thermal susceptibility as a function of the
temperature for different values of the magnetic field $B$. }
\label{suscep}
\end{figure}

Finally,  in Fig.~\ref{tcXB} we present the results for the pseudocritical $T_{\rm pc}$ 
temperature as a function of the magnetic field. It is clearly seen in this figure that
the pseudocritical temperature decreases as $B$ increases in (qualitative) 
agreement with the lattice results of Refs.~\cite{lattice1,lattice2}.

\begin{figure}[htb]
\centerline{ \epsfig{file=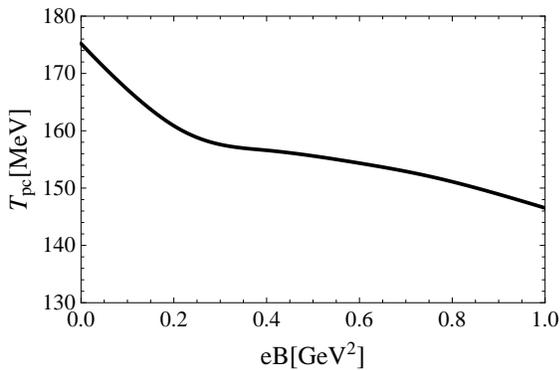,width=0.85\linewidth,angle=0}}
\caption{The pseudo critical temperature for the chiral transition of magnetized quark 
matter as a function of the magnetic field.}
\label{tcXB}
\end{figure}

The phase diagram shown in Fig.~\ref{tcXB} reproduces the lattice results qualitatively.
The figure shows a $T_{pc}$ dependence with $B$ that starts linearly at small $B$, 
with almost no curvature, while lattice results have almost vanishing linear term and 
a noticeable curvature. One may view this as a consequence of the rather simple form
used for the running of $G(B,T)$. However, it is important to note that our ansatz 
avoids the undesired ``turn over" effect seen in NJL models with a $B$-independent coupling,
in which at intermediate values of $B$, after an initial decrease, $T_{pc}$ starts to 
{\em increase} with $B$ - see e.g. Refs.~ \cite{debora,bruno}. 
A better quantitative agreement with the lattice phase diagram can certainly be obtained 
using a $G(B,T)$ with more fitting parameters.

\section{Conclusions and Perspectives}
\label{conclusions}

In this work we have considered the two flavor NJL model for hot and magnetized quark 
matter within the MFA. With the aim of understanding discrepancies between effective model
predictions and recent lattice results regarding the behavior of the chiral transition 
pseudocritical temperature as a function of $B$, we have examined the effect of introducing 
a running coupling $G$ motivated by asymptotic freedom.  Effective quark theories as the 
NJL model can be motivated by QCD by integrating out gluonic degrees of freedom. 
Although some features of confinement can be enforced by means of extending the model with 
the Polyakov loop, the running with energy scales of the effective 
coupling, as e.g. due to asymptotic freedom, is lost. In general, a decrease of $G$ with 
the temperature does not modify the qualitative features regarding the chiral phase 
transition~\cite{bernard}. However, as we have discussed, the same does 
not seem to be the case in the presence of a strong magnetic field when the formation of a 
quark condensate $\langle {\bar \psi}_ f \psi_f \rangle$ is enhanced by the magnetic 
catalysis effect. Generically, the quark mass gap at zero temperature sets the pseudocritical 
temperature at which chiral symmetry is (partially) restored. Since the effective quark mass 
$M$ in the NJL model, given by the gap equation $M \sim - G \langle {\bar \psi}_ f \psi_f \rangle$, 
increases with $B$, one would intuitively expect that $T_{\rm pc}$ should also increase; 
but this is not confirmed by lattice simulations. Here, we have shown that one way to conciliate 
the results is to suppose that $G$ decreases with $B$ and $T$ so as to moderate the increase 
of $M$ and, in turn, to the decrease of $T_{\rm pc}$ with $B$ in qualitative agreement with
the $T_{\rm pc}(B)$ obtained by the lattice simulations of Refs.~\cite{lattice1,lattice2}.

As future work one can improve the ansatz so as to get a better quantitative agreement with
the lattice results. Next, the model could be used in a plethora of situations in order to
analyze the behavior of physical quantities such as decay constants, meson-quark couplings,
among others which can be tested in lattice simulations. One could also extend the calculations
to the finite chemical potential domain, which currently is inaccessible to lattice 
calculations and investigate eventual additional running of $G$ with $\mu$. This is 
particularly interesting as a recent study has shown~\cite{preis} that the coexistence 
chemical potential decreases with $B$. Such a study is important in the context of the 
physics of magnetars.

Also, it would be interesting to further investigate why the results for $T_{\rm pc}(B)$ 
obtained  with $G(B,T)$ and $T_0(B)$ in the NJL and entangled PNJL models, respectively, are in 
qualitative agreement with the QCD lattice results while those obtained with the QM and 
PQM models are not. Another interesting new perspective is the investigation of the effect of
a $G(B,T)$ in the generation of spin-one condensates~\cite{Ferrer:2013noa}. Such spin-one
condensates give rise to a dynamical anomalous magnetic moment for the fermions and
lead to an increased critical temperature for chiral symmetry restoration for $B$ and 
$T$ independent NJL coupling.

Concluding, our assumption of the decrease of the effective four-quark coupling with $B$ 
and $T$ mimicking asymptotic freedom in QCD~\cite{miransky} represents a concrete implementation 
of the back reaction of the sea quarks and confirms its potential importance on explaining 
the inverse catalysis as stressed in the recent literature~\cite{endrodi,kojo}. In this respect,
a first principles nonperturbative framework to study the effects of a magnetic field in 
the continuum is provided by the Dyson-Schwinger equations at finite temperature and 
density~\cite{Roberts:2000aa}. The coupling of the magnetic field to the sea quarks is 
naturally taken into account via the coupled system of integral equations for the quark 
and gluon propagators. Very recently, this approach was used to study fermion mass generation 
in the presence of an external magnetic field in QED~\cite{{Ayala:2006sv},{Rojas:2008sg}} 
and QCD~\cite{kojo,{Watson:2013ghq},{Mueller:2014tea}}. The extension of such studies to 
finite temperatures would shadow light on the problem of inverse magnetic catalysis beyond 
the framework of effective chiral models.

\section*{Acknowledgments}
We thank Gergely Endrodi for discussions and also for providing the lattice data 
of the up and down quark condensates. We also thank Eduardo Fraga for bringing 
Ref.~\cite{bruno} to our attention. This work was partially supported by CNPq, 
FAPEMIG (R.L.S.F), FAPESP (G.K.),  FAPESC (M.B.P.) and CAPES (K.P.G). 
R.L.S.F. would like to thank J. Noronha and R.O. Ramos for discussions on 
related matters. 


\end{document}